\documentclass[12pt]{iopart}
\usepackage{graphicx,epsfig}

\begin{document}

\title{On virialization with dark energy}

\author{Irit Maor\dag\
        and Ofer Lahav\ddag}

\address{\dag\ Institute of Astronomy,
    University of Cambridge, Madingley Road,
    Cambridge CB3 0WA, UK~~
    {\it{i.maor@damtp.cam.ac.uk}}}
\address{\ddag\ Department of Physics and Astronomy,
    University College London,
    Gower Street, London WC1E 6BT, UK~~
    {\it{lahav@star.ucl.ac.uk}}}

\begin{abstract}
We review the inclusion of dark energy into the formalism of
spherical collapse, and the virialization of a two-component
system, made of matter and dark energy. We compare two approaches
in previous studies. The first assumes that only the matter
component virializes, e.g. as in the case of a classic
cosmological constant. The second approach allows the full system
to virialize as a whole. We show that the two approaches give
fundamentally different results for the final state of the system.
This might be a signature discriminating between the classic
cosmological constant which cannot virialize and a dynamical dark
energy mimicking a cosmological constant. This signature is
{\it{independent}} of the measured value of the equation of state.
An additional issue which we address is energy non-conservation of
the system, which originates from the homogeneity assumption for
the dark energy. We propose a way to take this energy loss into
account.
\end{abstract}

\maketitle

\section{Introduction}
\label{intro}

The top hat spherical collapse formalism dates back to Gunn and
Gott \cite{gg}. In addition to its beautiful simplicity, it has
proven to be a powerful tool for understanding and analysing the
growth of inhomogeneities and bound systems in the Universe. It
describes how a small spherical patch of homogeneous overdensity
decouples from the expansion of the Universe, slows down, and
eventually turns around and collapses. One assumes that the
collapse is not completed into a singularity, but that the system
eventually virializes and stabilizes at a finite size. The
definition of the moment of virialization depends on energy
considerations. The top hat spherical collapse is incorporated,
for example, in the Press-Schecter \cite{ps} formalism. It is
therefore widely used in present day interpretation of data sets.

For an Einstein-de Sitter Universe (EdS), i.e.~a Universe with
$\Omega_m=1$ and $\Omega_{\Lambda} = 0$ that is composed strictly
of non relativistic dust, there is an analytical solution for the
spherical collapse. The ratio of the final, virialized radius to
the maximal size at turnaround of the bound object is
$R_{vir}/R_{ta}=\frac{1}{2}$. The situation becomes more
complicated and parameter dependent when one considers a component
of dark energy. This was a subject of numerous studies
\cite{llpr,ws,is,wk,bw,zg,mb}. Lahav {\it et al} (LLPR)
\cite{llpr} generalized the formalism to a Universe composed of
ordinary matter and a cosmological constant. In this case the
cosmological constant is `passive', and only the matter
virializes. This leads to $R_{vir}/R_{ta}< \frac{1}{2}$. This
scenario also corresponds to the dynamics implemented in $N$-body
simulations for the concordance $\Lambda CDM$ model, i.e. the
cosmological constant only affects the time evolution of the scale
factor of the background Universe. Wang and Steinhardt (WS)
\cite{ws} included quintessence with a constant or slowly varying
equation of state. Battye and Weller (BW) \cite{bw} included
quintessence in a different manner to WS, taking into account its
pressure. Mota and Van de Bruck  (MB) \cite{mb} considered
spherical collapse for different potentials of the quintessence
field, and checked what happens when one relaxes the common
assumption that the
quintessence field does not cluster on the relevant scales. \\

In this work we wish to review the inclusion of a cosmological
constant and quintessence into the formalism of spherical
collapse. Adding dark energy creates a system with two components
- the matter and the additional fluid. Most existing works
\cite{llpr,ws,bw,zg} look at the virialization of the matter
component (luminous and dark), which feels an additional potential
due to the presence of the dark energy. With this procedure, one
implicitly assumes that the dark energy either does not virialize,
or does so separately from the matter component. MB on the other
hand, included the additional fluid in the virialization
equations, the assumption here being that all of the system's
components virialize together. However, they did not remark either
on the difference in physical understanding of the system between
their approach and the one common in the literature, or on the
case of the cosmological constant. Our aim here is to critically
contrast the two approaches - the assumption that the dark matter
component virializes separately (as in LLPR, WS, and BW), and the
assumption that the whole system virializes as a whole (as in MB).
We wish to consider the meaning of including or not including the
additional fluid into the virialization, and point out a few
puzzles.

A second issue which we will address here is the use of energy
conservation in order to find the condition of virialization.
Assuming that the quintessence field does not collapse with the
mass perturbation but stays homogeneous as the background means
that the system must lose energy as it collapses. Yet, energy
conservation between turnaround and virialization is assumed. This
inconsistency is for quintessence fields with equation of state
$w\neq -1$. Energy is conserved with a cosmological constant
($w=-1$), for reasons that will be discussed later on. We will
propose a way to take into account this energy loss for
quintessence, and introduce a correction to the equation that
defines the final virialized radius of the system. \\

The inclusion of dark energy in the virialization process changes
the results in a fundamental manner. As we will show, the ratio of
final to maximal size of the spherical perturbation is
{\it{larger}} if the dark energy is part of the virialization.

While the results we will show are of the cosmological constant or
quintessence with a constant $w$, the methods we use are
applicable to a time dependent equation of state, as well as to
models in which quintessence is coupled to matter \cite{qc}. We
limit the discussion here to $w\geq-1$. While an equation of state
which is more negative than $-1$ is observationally interesting,
the physical interpretation of it is unclear, and beyond the scope
of this work. We assume throughout that the background is
described by a flat, FRW metric, with two energy components - the
matter and the dark energy.

The paper is organized as follows. In section \ref{1cs} we give
the general picture of how one calculates the point of
virialization of a single component system, and define the
relevant fundamental quantities. In section \ref{2cs} we consider
the case of a two component system. Section \ref{cq} reviews the
case of a clustered quintessence. In section \ref{g} we examine
the transition from clustered to homogeneous quintessence and in
section \ref{cc} we examine the transition toward $w=-1$. We
summarize and conclude in section \ref{conclusions}.

\section{Virialization of a single component system}
\label{1cs} The spherical collapse provides a mathematical
description of how an initial inhomogeneity decouples from the
general evolution of the Universe, and expands in a slower
fashion, until it reaches the point of turnaround and collapses on
itself. The mathematical solution gives a point singularity as the
final state. Physically though, we know that objects go through a
virialization process, and stabilize towards a finite size.

Since virialization is not `built in' into the spherical collapse
model (see though \cite{pad}), the common practice is to
{\it{define}} the virialization radius as the radius at which the
virial theorem holds, and the kinetic energy $T$ is related to the
potential energy $U$ by $T_{vir}=\frac{1}{2}(R ~\partial
U/\partial R)_{vir}$. Using energy conservation between
virialization and turnaround (where $T_{ta}=0$) gives
\begin{eqnarray}
 \left[ U+\frac{R}{2}\frac{\partial U}{\partial R}
    \right]_{vir} & = & U_{ta} \label{ec} ~.
\end{eqnarray}
Equation (\ref{ec}) defines $R_{vir}$. Thus in order to calculate
the final size of a bound object, we need to know how to calculate
the potential energy of the spherical perturbation, and to use
energy conservation between turnaround and the time of
virialization. We discuss later the case where energy is not
conserved, and how to account for it. For an EdS Universe,
$U=-\frac{3}{5}GM/R$ ($M$ is the conserved mass within the
spherical perturbation) and $T_{vir}=-\frac{1}{2}U$, so the ratio
of final to maximal
radii of the system is $x=R_{vir}/R_{ta}=\frac{1}{2}$. \\

The virial equation, which at equilibrium gives the above results,
is usually derived from the Euler Equation for particles. It is
worth noting here that one can derive the virial equation for a
cosmological fluid. The starting point is the continuity equation
of a perfect fluid (derived from $T^{0\nu}~_{;\nu}=0$), with
equation of state (the ratio of pressure to energy density)
$w=p/\rho$:
\begin{eqnarray}
 \dot\rho+3\left( 1+w \right) \frac{\dot r}{r}\rho &=& 0 ~.
\label{vir1}
\end{eqnarray}
Multiplying equation (\ref{vir1}) by $r^2$, taking the time
derivative and integrating over a sphere of radius $R$ gives
\begin{eqnarray}
 \int\dot G dV+\frac{1}{2}\left(1+3w \right)\left[
   \int\rho\dot r^2 dV+\int r\frac{d}{dt}\left(\rho \dot r
   \right)dV \right] &=& 0 ~,
\label{vir2}
\end{eqnarray}
where $G=(d/dt)\left(\frac{1}{2} \rho r^2\right)$. In the
classical analogy, $\dot G=\ddot I$ is the second derivative of
the inertia tensor. In a state of equilibrium, $\dot G=0$. The
quantity $\int\rho\dot r^2 dV$ is twice the kinetic energy, and
$\int r(d/dt)\left(\rho \dot r \right)dV $ is $R~ \partial
U/\partial R$. As equation (\ref{vir2}) shows, the value of $w$
factorizes out when one is looking for the equilibrium condition.
In the case where the fluid does not conserve energy, the right
hand side of of equation (\ref{vir1}) will be equal to some
function $\Gamma$. In that case, the virial equation (\ref{vir2})
will have an additional surface term. In equilibrium, the surface
term and
$\dot G$ should vanish. \\
This is the non-relativistic version of the scalar virial theorem.
Hence, it is not applicable for a fluid with a relativistic
equation of state, $w\rightarrow \frac{1}{3}$. It can be shown
that when writing the relativistic version, the energy of the
radiation field drops out of the virial equation \cite{ll}.

\section{A two - component system}
\label{2cs}

When adding a component to the system, there are three questions
to be asked - {\bf{(a)}} how does the potential induced by the new
component affect the system? {\bf{(b)}} does the new component
participate in the virialization? and {\bf{(c)}} does the new
component cluster, or does it stay homogeneous? These questions
should be addressed in the framework of a fundamental theory for
dark energy. Here we work out the consequences of virialization
and clustering of dark energy, if they do take place. We
shall try to address each of these questions separately. \\
\vspace{0.50cm}\\
{\bf{(a)}} In the case where the new component does not cluster or
virialize, its sole effect is contributing to the potential energy
of component 1. LLPR \cite{llpr} calculated this contribution to
the potential energy using the Tolman-Bondi equation. Their result
was generalized for quintessence by WS \cite{ws} and, in a
different manner, by BW \cite{bw}. We follow LLPR and WS in our
numerical calculations.
\vspace{0.50cm}\\
{\bf{(b)}} Any energy component with non vanishing kinetic energy
is capable of virializing, but the question is: on what time
scale? If one imagines that the full system virializes, then the
virial theorem should relate the {\it{full}} kinetic and potential
energies of the system,
\begin{eqnarray}
 U &=& U_{11}+U_{12}+U_{21}+U_{22}=
   \frac{1}{2}\int\left(\rho_1+\rho_2\right)
   \left(\Phi_1+\Phi_2 \right)dV \label{uf} ~,
\end{eqnarray}
where the potential $\Phi_x$ induced by each energy component in a
spherical homogeneous configuration is
\begin{eqnarray}
 \Phi_x(r) &=& -2\pi G (1+3w_x)\rho_x\left(R^2-\frac{r^2}{3}
    \right)~.
\end{eqnarray}
The kinetic energy at virialization is then
\begin{eqnarray}
 T_{tot} &=& \frac{1}{2}R\frac{\partial}{\partial R}
 \left(U_{11}+U_{12}+U_{21}+U_{22}\right) ~.
\label{vf}
\end{eqnarray}
The expression above is the full potential energy of the system.
As we will show, the addition of these new terms to the virial
theorem makes a fundamental difference in the final state of the
system, so the question of whether dark energy participates in
the virialization is crucial. \\
\vspace{0.50cm}\\
{\bf{(c)}} Every positive energy component other than the
cosmological constant is capable of clustering. Even though
Caldwell {\it{et al}} \cite{c} have shown that quintessence cannot
be perfectly smooth, it is assumed that the clustering is
negligible on scales less than $100~Mpc$. It is therefore common
practice to keep the quintessence homogeneous during the evolution
of the system. The effects of relaxing this assumption were
explored in MB. We will consider both cases here.
The continuity equation for a Q component which is kept
homogeneous is
\begin{eqnarray}
  \dot\rho_{Qc}+3(1+w)\frac{\dot a}{a}\rho_{Qc} & = & 0 ~,
\end{eqnarray}
and for clustering Q is
\begin{eqnarray}
  \dot\rho_{Qc}+3(1+w)\frac{\dot r}{r}\rho_{Qc} & = & 0
\end{eqnarray}
($a$ and $r$ are the global and local scale factors,
respectively). One can ask what happens if one slowly `turns on'
and enables the possibility of clustering for the Q component. To
enable a slow continuous `turn on' of the clustering, one can
write
\begin{eqnarray}
 && \dot\rho_{Qc}+3(1+w)\left(\frac{\dot r}{r}\right)\rho_{Qc} = \gamma \Gamma
    \label{13}\\
 && \Gamma  =  3(1+w)\left(\frac{\dot r}{r}-
 \frac{\dot a}{a}\right)\rho_{Qc} \\
 && 0\leq \gamma \leq  1 ~,
\end{eqnarray}
where $\rho_{Qc}$ is the dark energy's density within the cluster.
The notation $\Gamma$ follows MB. $\gamma$ is the `clustering
parameter', $\gamma=0$ gives clustering behaviour and $\gamma=1$
gives homogeneous behaviour. In the case of $\gamma=1$, the dark
energy inside the spherical region and the background dark energy
$\rho_Q$, behave in similar ways: $\rho_{Qc}=\rho_Q$. A point to
bear in mind is that for the case of homogeneous quintessence, the
system does not conserve energy as it collapses from turnaround to
virialized
state. \\

Putting all this together, the equations governing the dynamics of
the spherical perturbation are
\begin{eqnarray}
\left(\frac{\ddot r}{r} \right) & = &-\frac{4\pi G}{3}
    \left(\frac{}{}\rho_{mc}+\left(1+3w \right)\rho_{Qc}\right) \label{qr2} \\
 \dot\rho_{mc} & + & 3\left(\frac{\dot r}{r}\right)\rho_{mc} = 0\\
 \dot \rho_{Qc} & + & 3\left(1+w \right)\left(\frac{\dot r}{r}\right)
    \rho_{Qc} = \gamma \Gamma \label{qc} ~.
\end{eqnarray}

\begin{figure}
  \begin{center}
    \epsfig{file=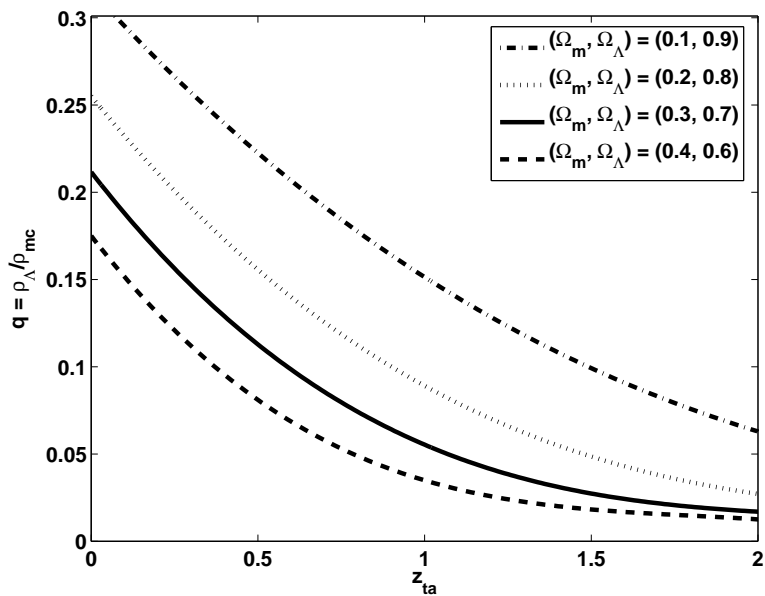,height=80mm}
  \end{center}
  \caption{$q=\rho_{\Lambda}/\rho_{mc}$ as a function of
  the turnaround redshift $z_{ta}$, for various
  values of $\Omega_m$ and $\Omega_{\Lambda}$.
  \label{qz}}
\end{figure}

Our results are going to be presented as a function of $q$, which
is defined as the ratio of the system's dark energy to matter's
densities at the time of turnaround,
$q(z_{ta})\equiv\rho_{Qc}(z_{ta})/\rho_{mc}(z_{ta})$. The system's
matter density $\rho_{mc}$ at turnaround is
$\rho_{mc}(z_{ta})=\zeta(z_{ta})\rho_m(z=0)(1+z_{ta})^3$,
expressed in terms of the background matter density
$\rho_m(z_{ta})$ and the density contrast at turnaround,
$\zeta(z_{ta})$. In order to estimate which values of $q$ are of
interest,we plotted, in figure \ref{qz}, the dependence of $q$ on
the turnaround redshift $z_{ta}$, for various values of $\Omega_m$
and $\Omega_{\Lambda}$. As can be seen, $q$ takes typical values
no larger than $0.3$.

\section{A clustered quintessence}
\label{cq}

In the case of fully clustering quintessence, $\gamma=0$, the
quintessence field responds to the infall in the same way as
matter, the only difference being its equation of state which
dictates a different energy conservation (in the general
relativity sense). However, energy is conserved, and since the
quintessence is active in the dynamics of the system, it is quite
reasonable to imagine that it takes part in the virialization. We
therefore present here the calculation assuming the whole system
virializes, matter and dark energy. \\

Following equation (\ref{uf}), the potential energy of the full
system is
\begin{eqnarray}
  U &=& -\frac{3}{5}\frac{GM^2}{R}-(2+3w)\frac{4\pi G}{5}M\rho_{Qc} R^2
        -(1+3w)\frac{16\pi^2 G}{15}\rho_{Qc}^2R^5 ~. \label{qec}
\end{eqnarray}

Once the potential energy has been calculated, virialization is
found with the use of equation (\ref{ec}). Expressing it in term
of $q=\rho_{Qc}/\rho_{mc}$ at turnaround and $x=R_{vir}/R_{ta}$
gives
\begin{eqnarray}
  &&\left[1+(2+3w)q+(1+3w)q^2\right]x \nonumber \\
  &&-\frac{1}{2}(2+3w)(1-3w)qx^{-3w}-
  \frac{1}{2}(1+3w)(1-6w)q^2x^{-6} =  \frac{1}{2} ~.
  \label{gen0}
\end{eqnarray}
Equation (\ref{gen0}) is valid under the assumption that the whole
system virializes. \\

If, on the other hand, only the matter virializes, then the
equation defining $x$ is
\begin{eqnarray}
 &&\left(1+q \right)x-\frac{q}{2}\left(1-3w \right)x^{-3w}
 =\frac{1}{2} ~.
 \label{ws0}
\end{eqnarray}

\begin{figure}
  \begin{center}
    \epsfig{file=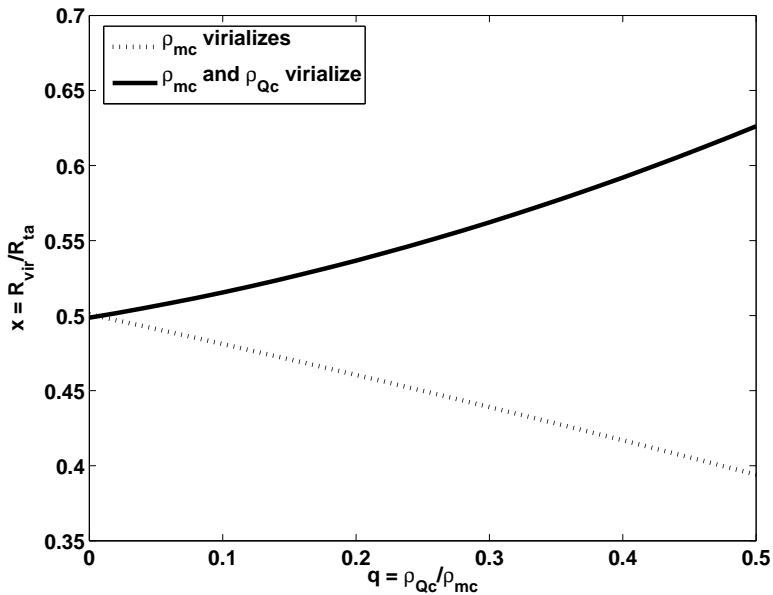,height=80mm}
  \end{center}
  \caption{The ratio of final to turnaround radii as a function
  of $q=\rho_{Qc}/\rho_{mc}$ at turnaround, for quintessence with
  a constant equation of state, $w=-0.8$, which fully clusters
  ($\gamma=0$). The dotted line is the ratio in the case where
  only matter virializes. The solid line
  is the ratio when the whole system,
  including the dark energy component, has virialized.
  \label{wc08}}
\end{figure}
In figure \ref{wc08} we show $x$ as a function of $q$, for a fluid
with $w=-0.8$ that fully clusters, $\gamma=0$. The dotted line is
the ratio in the case where only matter virializes, equation
(\ref{ws0}). The solid line is the ratio when the whole system,
including the dark energy component, have virialized, equation
(\ref{gen0}). As can be seen, there is a fundamental difference of
the solutions: if only the matter virializes then the final ratio
is smaller than the EdS $\frac{1}{2}$ value, while when the whole
system virializes, the final ratio is larger than $\frac{1}{2}$.

\section{Turning off the clustering}
\label{g}

It is the usual practice to neglect spatial perturbations of the
quintessence field, and to keep it homogeneous \cite{c}. With our
generalized notation of the `clustering parameter' $\gamma$, one
can also allow a small but non-zero amount of clustering for the
quintessence. \\
For any $\gamma\neq 0$, the quintessence field within the system
does not conserve energy. As equation (\ref{ec}) assumes energy
conservation, the problem with not allowing the quintessence to
fully cluster is how to find the radius of virialization. We will
now propose a correction to equation (\ref{ec}), that will take
into account the loss of energy.

We denote the potential energy at turnaround as $U_{ta}$, and at
virialization as $U_{vir}$. We also define a function $\tilde U$
as the system's potential energy {\it had} it conserved energy.
Thus by construction the energy that the system lost is
\begin{eqnarray}
 \Delta U & \equiv & \tilde U-U ~.
\end{eqnarray}
Equation (\ref{ec}) which describes energy conservation between
turnaround and virialization now needs to be corrected. Accounting
for the lost energy gives
\begin{eqnarray}
\left[ U+\frac{R}{2}\frac{\partial U}{\partial R}
    \right]_{vir}+\Delta U_{vir}  =
    \left[\tilde U+\frac{R}{2}\frac{\partial U}{\partial
    R}\right]_{vir}
    = U_{ta} ~. \label{new}
\end{eqnarray}
We are now set to calculate $\Delta U$. \\

Looking at equation (\ref{qec}), one can treat $U$ as
$U(\rho_x,R)$ ($\rho_x$ being the various energy density
components). In order to calculate $\tilde U(\rho_x,R)$, one needs
to replace $\rho_x$ with $\tilde \rho_x$ in the expression for
$U$, which has $\gamma=0$ and conserves energy. The continuity
equation for $\tilde \rho_x$ is then
\begin{eqnarray}
    \dot{\tilde \rho_x}+3\left(\frac{\dot R}{R} \right)
        \left(1+w_x\right)\tilde\rho_x=0 ~, \label{tilde}
\end{eqnarray}
and we impose boundary conditions such that
$\tilde\rho_x(a_{ta})=\rho_x(a_{ta})$. \\
For a constant equation of state, this gives
\begin{eqnarray}
    \tilde\rho_x & = & \tilde \rho_x(a_{ta})
        \left(\frac{R_{ta}}{R_{vir}} \right)^{3(1+w_x)}
    =\rho_x(a_{ta})
        \left(\frac{R_{ta}}{R_{vir}} \right)^{3(1+w_x)} ~.
\end{eqnarray}
For a time dependent $w$ one needs to use the integral
expression for $\tilde \rho$. \\
We therefore have
\begin{eqnarray}
    \tilde U(\rho_x,R) & = & U(\tilde \rho_x, R) ~.
\end{eqnarray}

Equation (\ref{new}) is now a function of $R_{vir}$ and values
determined at turnaround time (such as $U_{ta}$ and
$\rho_x(a_{ta})$), and defines $R_{vir}$ in the same manner as
equation (\ref{ec}) did. With the definitions of $q$, $x$,
$y=(a_{vir}/a_{ta})^{1+w}$ and $p=x/y$, the final form of equation
(\ref{ec}) for a quintessence with a general value of $\gamma$ is
then
\begin{eqnarray}
  &&
  \frac{q^2}{2}\left(1+3w\right)
  \left[\frac{}{}1+6w-6\gamma\left(1+w\right)\right]
  \left[\frac{}{}\left(1-\gamma \right)x^{-3w}+
    \gamma p^3\right]^2
  \nonumber \\
  &+&
  \frac{q}{2}\left(2+3w\right)
   \left[\frac{}{}1+3w-3\gamma\left(1+w\right)\right]
   \left[\frac{}{}\left(1-\gamma \right)x^{-3w}+\gamma p^3 \right]
  \nonumber \\
  &+&
  \left[\frac{}{}1+\left(2+3w \right)q+\left(1+3w\right)q^2 \right]x
  -\left(2+3w \right)qx^{-3w}- \left(1+3w\right)q^2x^{-6w}
  \nonumber \\
  &=&  \frac{1}{2} ~.
\label{general}
\end{eqnarray}
For the case (common in literature) of a completely homogeneous
quintessence, $\gamma=1$, the virialization condition
(\ref{general}) is reduced to
\begin{eqnarray}
 && \left[1+(2+3w)q+(1+3w)q^2\right]x- \nonumber \\
 && (2+3w)q\left(p^3+x^{-3w}\right)-
    (1+3w)q^2\left(\frac{5}{2}p^6+x^{-6w}\right)=  \frac{1}{2} ~.
 \label{gen1}
\end{eqnarray}
\\

The equation for $x$ when only the matter virializes does not need
to be corrected for energy conservation, as it counts only the
energy associated with the matter. Its general form is
\begin{eqnarray}
 && \left(1+q\right)x-\frac{q}{2}
    \left[\frac{}{}1-3w+3\gamma\left(1+w\right)\right]
    \left[\left(1-\gamma \right) x^{-3w}+\gamma p^3 \right]
 =\frac{1}{2} ~,
 \label{wsg}
\end{eqnarray}
and for $\gamma=1$ it reproduces the solution of WS,
\begin{eqnarray}
 && \left(1+q\right)x-2qp^3 =\frac{1}{2} ~,
 \label{ws1}
\end{eqnarray}
\\
which is a generalization of LLPR's results  (to be discussed
later on, see equation (\ref{llpr})).

One should consider which of the solutions is more plausible.
Ultimately the choice between the two solutions should be dictated
by the theory with which the dark energy is modelled. As we are
looking at an effective description of the dark energy as a
perfect fluid and not with a fundamental theory, this information
is lost.

\begin{figure}
  \begin{center}
    \epsfig{file=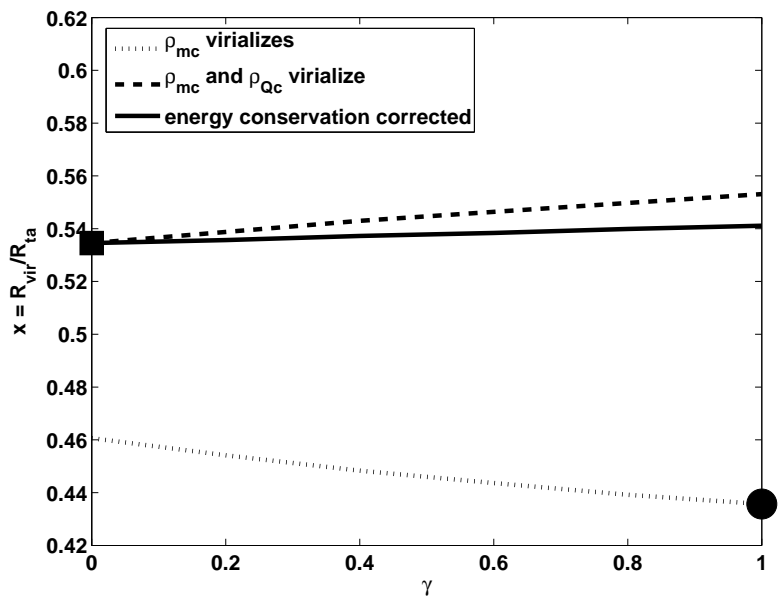,height=80mm}
  \end{center}
  \caption{$R_{vir}/R_{ta}$ as a function of $\gamma$, for
  $w=-0.8$ and $q=0.2$. $\gamma=0$ describes the case of a
  fully clustering $Q$
  field, and $\gamma=1$ is the case of a homogeneous $Q$,
  allowing only the matter component to cluster. For
  $\gamma=0$, taking the dark energy into the virialization is
  highly plausible, (see square on left).
  If one assumes that only the matter component virializes
  for $\gamma=1$ (see circle on right), it is unclear how to
  extrapolate in a smooth way between the two cases.
  This will produce a discontinuity in the transition
  from the `clustering' to the `non-clustering' behaviour.
  \label{gamma}}
\end{figure}

For the case of $\gamma=1$, when one keeps the evolution of the
dark energy in the system identical to that of the background, it
is reasonable to assume that it does not participate in the local
processes that lead to virialization. This gives credibility to
the solution of equation (\ref{ws1}), allowing only the matter to
virialize. However, this presents a question of continuity,
presented in figure \ref{gamma}. The figure shows the solutions of
$x$ as a function of $\gamma$, with fixed $w=-0.8$ and $q=0.2$.
The circle on the right is the WS's result when the quintessence
is kept completely homogeneous. The square on the left is the
result when both the matter and the quintessence virialize, for
the fully clustering case. The `clustering parameter' allows us to
think of a continuous transition between the two cases. One would
expect the transition in the behavior of the system along $\gamma$
to be smooth. Allowing the dark energy to virialize for the
clustering case, $\gamma=0$, and keeping it out of the
virialization process when $\gamma=1$, raises the question of how
one should extrapolate smoothly between the two cases. As figure
\ref{gamma} suggests, there will be a
discontinuity. \\
\begin{figure}
  \begin{center}
    \epsfig{file=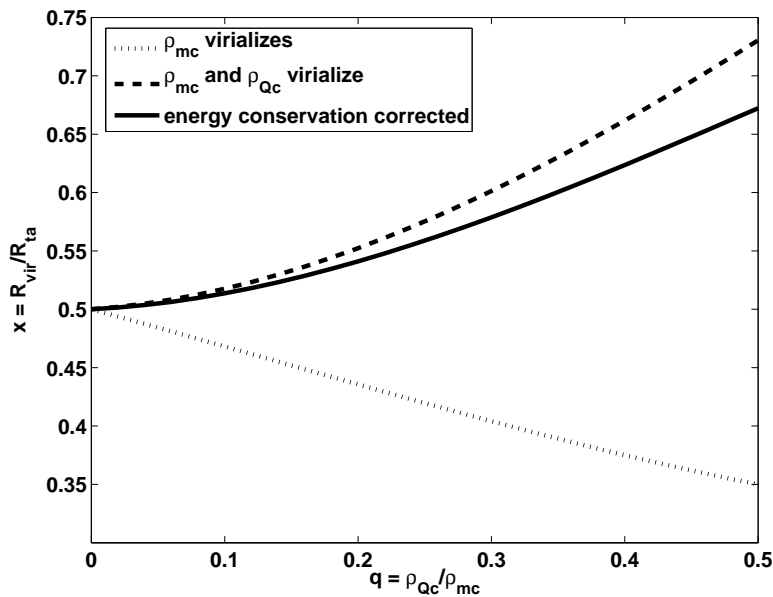,height=80mm}
  \end{center}
  \caption{The ratio of final to turnaround radii as a function
  of $q=\rho_{Qc}/\rho_{mc}$ at turnaround, for quintessence with
  a constant equation of state, $w=-0.8$, which stays homogeneous
  ($\gamma=1$). The dotted line follows
  WS's calculation, assuming only the matter component virializes.
  The dashed line is the ratio when the whole system,
  including the dark energy component, has virialized. The solid
  line takes into account the loss of energy between turnaround
  and virialization.
  \label{r08}}
\end{figure}

In figure \ref{r08} we compare the solution of equation
(\ref{ws1}) and (\ref{gen1}) for $w=-0.8$ and $\gamma=1$, and show
the effect of the energy correction. As can be seen, taking into
account the loss of energy produces a small quantitative
correction, but keeps the general feature of enlarging the final
size of the system if the dark energy is allowed to virialize.

\section{The limit of a cosmological constant}
\label{cc}

Equations (\ref{general}) and (\ref{wsg}) are valid for any
constant $w$. As $w$ approaches $-1$, we get that $p\rightarrow
x$, and the dependency on $\gamma$ vanishes. The reason that
$\gamma$ plays no role in the limit of $w\rightarrow -1$ is that
the question whether such a fluid is allowed to cluster
($\gamma=0$) or not ($\gamma=1$) is rather abstract. It stays
homogeneous in any case, because of its equation of state,
$w_{\Lambda}=-1$ (which leads to
$\Gamma=0$). Accordingly, energy is automatically conserved. \\

In that limit, equation (\ref{general}) which assumes that the
whole system to virialize, is simplified into
\begin{eqnarray}
7 q^2 x^6 + 2 q x^3 + \left(1-q-2q^2\right)x & = & \frac{1}{2} ~.
\label{gencc}
\end{eqnarray}

Taking the same limit for equation (\ref{wsg}) yields the familiar
result of LLPR:
\begin{eqnarray}
  \left(1+q\right)x-2qx^3 & = & \frac{1}{2} ~~~~~(LLPR) ~,
\label{llpr}
\end{eqnarray}
which is valid under the assumption that the matter component
alone virializes \footnote{One can look at a test particle feeling
an inverse square force and an additional repulsive $\Lambda$
force. Consider two possible orbits of the particle: one circular,
and one in which its kinetic energy can vanish. The ratio of the
circular radius to the radius of zero kinetic energy
(`turnaround') is described exactly by equation (\ref{llpr}). This
assumes that the test particle does not contribute to the forces
of the system. We thanks Martin Rees for pointing out this
similarity.} (notice that our definition of $q$ differs by a
factor of $2$ from the definition of $\eta$ of LLPR,
$q=\frac{1}{2}\eta$). \\

Again, we wish to consider the plausibility of the two solutions.
If one considers the cosmological constant as a true constant of
Nature, $\rho_{\Lambda}=\Lambda/(8 \pi G)$, it is hard to imagine
it participating in the dynamics that lead to virialization, as it
is a true constant. In this case, one could categorically say that
the right procedure is to look at the virialization of the matter
fluid only, and follow LLPR's solution, equation (\ref{llpr}). The
sole effect of the cosmological constant, then, is to modify the
potential that the matter fluid feels.

If, on the other hand, one considers the origin of a perfect fluid
with $w\approx -1$ as a special case of quintessence, which is
indistinguishable from a cosmological constant, it is reasonable
to expect continuity in the behaviour of the system as one slowly
changes the value of $w$ toward $-1$. In other words, if the
physical interpretation of the fluid with $w\approx -1$ is of a
dynamical field that {\it{mimics}} a constant, the idea of
including it in the dynamics of the system has a physical meaning.

\begin{figure}
  \begin{center}
    \epsfig{file=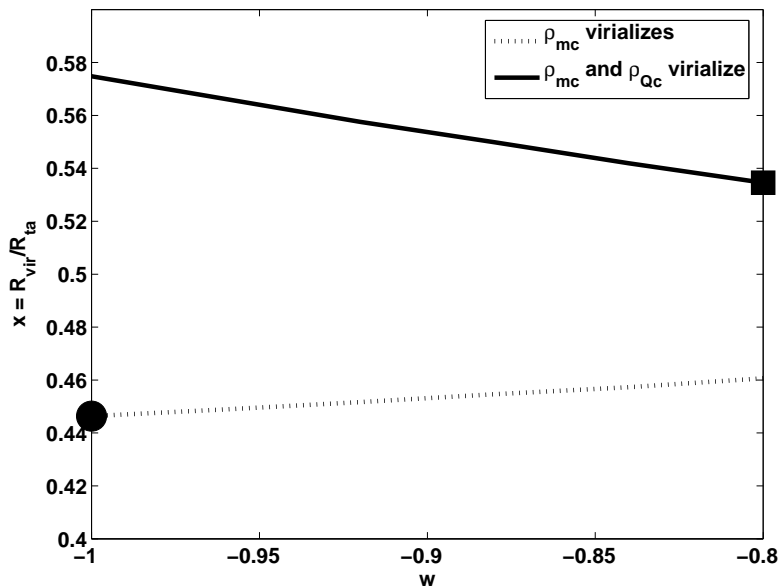,height=80mm}
  \end{center}
  \caption{$R_{vir}/R_{ta}$ as a function of $w$, for
  $q=0.2$ and $\gamma=0$. The dotted line is the ratio when
  the matter alone virializes, and the solid is for the case where
  the whole system virializes. The circle on the left is LLPR's
  solution for the cosmological constant. The square on the right
  is an example of a clustered quintessence, where we expect to
  take into account the whole system in the virialization. The
  figure suggests we should expect a discontinuity in the
  behaviour of quintessence fields and a true cosmological
  constant.
    \label{w0}}
\end{figure}

The result, then, is that we possibly have a signature
differentiating between a cosmological constant which is a true
constant, and something else which {\it{mimics}} a constant. This
point is shown in figure \ref{w0}. The figure shows $x$ as a
function of $w$, with $q=0.2$ and $\gamma=0$. The dotted line
follows the solution of equation (\ref{wsg}), with the matter
alone virializing. The circle on the left is LLPR's solution for
the cosmological constant. The solid line follows the solution of
equation (\ref{general}). The square on the right is an example of
a clustered quintessence, where we expect to take into account the
whole system in the virialization. As with figure \ref{gamma},
there is a suggested discontinuity, but here one can associate the
discontinuity with a clear physical meaning: a true cosmological
constant is not on the continuum of perfect fluids with general
$w$, as its physical behaviour is different. \\

An observational detection of virialized objects with
$R_{vir}>\frac{1}{2}R_{ta}$ would be a strong evidence against a
cosmological constant which is a true constant, regardless of the
measured value of the equation of state.

\section{Conclusions}
\label{conclusions}

In this work, we have reconsidered the inclusion of a dark energy
component into the formalism of spherical collapse. We compared
existing results (such as those of LLPR and WS) which implicitly
assume that only the dark matter virializes, to the case where the
whole system's energy is taken into account for virialization,
implying that the dark energy component also participates in the
process (MB). While previous studies allow the dark energy
component either to fully cluster or keep completely homogeneous,
we generalized and allowed a smooth transition between the two
cases. Additionally, we addressed the issue of energy
non-conservation when the dark energy is kept homogeneous.

Our main conclusions are:
\begin{itemize}
\item
In the case of a true cosmological constant only the matter
component virlializes and the LLPR solution is valid.

\item
If both components of the system virialize, two additional terms
to the potential energy appear. These are the self - energy of the
additional energy source, and its reaction to the presence of the
matter.
\item
The inclusion of these terms results in a fundamentally different
behaviour of the system. If only dark matter virializes, the final
size of the system is {\it{smaller}} than half of its maximal
size. When the whole system virializes, its final size is
{\it{bigger}} than half of its maximal size.
\item
It is hard to understand the physical meaning of a cosmological
constant `virializing', if it is a true constant. Accordingly,
observational evidence for $R_{vir}>\frac{1}{2}R_{ta}$ would be
strong evidence in favour of a dynamical field for the dark
energy, regardless of the measured value of the equation of state.
On the other hand, $R_{vir}<\frac{1}{2}R_{ta}$ is compatible with
both a true constant and a field mimicking the cosmological
constant.
\item
Keeping the dark energy component homogeneous implies that the
overdense region does not conserve energy. The exception here is
the case of the cosmological constant, for which the
non-clustering behaviour is exact and not an approximation. The
equation defining virialization needs to be corrected, in order to
account for the energy lost by the Q field between turnaround and
virialization. It should read
\begin{eqnarray}
    \left[\tilde U+\frac{R}{2}\frac{\partial U}{\partial
    R}\right]_{vir} & = & U_{ta} ~. \nonumber
\end{eqnarray}
This introduces a small quantitative correction.
\end{itemize}

Table \ref{table} gives a summary of the relevant solution for the
different cases that we considered in this work.
\begin{table}
  \begin{center}
    \begin{tabular}{||c||c|c||}
    \hline \hline
    &{\bf{$\rho_{mc}$ virializes}}
        &{\bf{$\rho_{mc}$ and $\rho_{Qc}$ virialize}} \\
    \hline \hline
    {\bf{general case}} & (\ref{wsg}) & (\ref{general}) \\
    \hline
    {\bf{$\gamma=0$}}   & (\ref{ws0}) & (\ref{gen0})    \\
    \hline
    {\bf{$\gamma=1$}}   & (\ref{ws1}) & (\ref{gen1})    \\
    \hline
    {\bf{$w\rightarrow -1$}} & (\ref{llpr})& (\ref{gencc})   \\
    \hline
    {\bf{Cosmological Constant}}& (\ref{llpr})& --   \\
    \hline \hline
    \end{tabular}
    \end{center}
    \caption{A summary of the relevant equations defining $x$ for
    the various cases that we considered.
    \label{table}}
\end{table}

Our work has consequences for the linear theory as well. As we
have not altered any of the equations governing the evolution of
the system, the linear equation of growth will not be altered
either. Nonetheless, reconsidering the energy budget changed the
time in which we perceive virialization to happen, and as a
consequence the linear contrast at virialization ($1.686$ for the
EdS case) will change. In practice though, the numerical change is
rather small. We find that for the cosmological constant, the
maximal deviation from the EdS value is a rise of about $3\%$.

A future work would be to incorporate the possible virialization
of dark energy into numerical simulations and into analyses of
observations. It would be particularly interesting to see how it
affects cluster abundances, and which approach provides a better
fit to the observations. Several works are pursuing such
directions \cite{additional}.

For various models of coupled quintessence \cite{qc}, it is very
likely that the dark energy component clusters and virializes. For
models in which the dark energy doesn't cluster, one could ask how
plausible the scenario of the dark energy participating in the
virialization is. Of course, should one argue that the dark energy
virializes and not just the dark matter component, a mechanism of
how it physically happens would be needed. An additional direction
to pursue is the actual mechanism of virialization, which at the
moment is still rather obscure. Understanding what the physical
process by which the system virializes is will hopefully give us a
clue as to whether we should include the dark energy or not.

\ack We would like to thank Jacob Bekenstein, Carsten van de
Bruck, Ramy Brustein, Uri Keshet, Donald Lynden-Bell, David Mota,
Amos Ori, Martin Rees and Jochen Weller for useful discussions. OL
acknowledges a PPARC Senior Research Fellowship. IM acknowledges
the support from the Leverhulme Quantitative Cosmology grant. Part
of this investigation was carried out while one of us (IM) was
visiting the Weizmann Institute for Science. IM wishes to thank
Micha Berkuz and Yosi Nir for their kind invitation and warm
hospitality.

\end{document}